# An extreme value statistics model of heterogeneous ice nucleation for quantifying the stability of supercooled aqueous systems

**Anthony N. Consiglio[1]\*, Yu Ouyang[1], Matthew J. Powell-Palm[2], Boris Rubinsky[1]**

[1]Department of Mechanical Engineering, University of California, Berkeley, CA, USA

[2]J. Mike Walker '66 Department of Mechanical Engineering, Texas A&M University, College Station, TX, USA

**\*Correspondence:**
Anthony N. Consiglio
aconsigilio4@berkeley.edu

**ABSTRACT**

The propensity of water to remain in a metastable liquid state at temperatures below its equilibrium melting point holds significant potential for cryopreserving biological material such as tissues and organs. The benefits conferred are a direct result of progressively reducing metabolic expenditure due to colder temperatures while simultaneously avoiding the irreversible damage caused by the crystallization of ice. Unfortunately, the freezing of water in bulk systems of clinical relevance is dominated by random heterogeneous nucleation initiated by uncharacterized trace impurities, and the marked unpredictability of this behavior has prevented implementation of supercooling outside of controlled laboratory settings and in volumes larger than a few milliliters. Here, we develop a statistical model that jointly captures both the inherent stochastic nature of nucleation using conventional Poisson statistics as well as the random variability of heterogeneous nucleation catalysis through bivariate extreme value statistics. Individually, these two classes of models cannot account for both the time-dependent nature of nucleation and the sample-to-sample variability associated with heterogeneous catalysis, and traditional extreme value models have only considered variation of the characteristic nucleation temperature. We conduct a series of constant cooling rate and isothermal nucleation experiments with physiological saline solutions and leverage the statistical model to evaluate the natural variability of kinetic and thermodynamic nucleation parameters. By quantifying freezing probability as a function of temperature, supercooled duration, and system volume, while accounting for nucleation site variability, this study also provides a basis for the rational design of stable supercooled biopreservation protocols.

## I. INTRODUCTION

Understanding how solid phases nucleate from the melt is a complex problem of wide-spread relevance. The nucleation of ice, in particular, governs the formation of clouds in our atmosphere[1], plays a key role in the survival of organisms living in extreme climates[2,3], and determines the quality of our refrigerated and frozen foods[4,5].

Ice nucleation also sets the current limit for the preservation of live saving organs for transplantation[6]. The clinical standard for *ex vivo* storage of organs involves placing an organ on ice (at roughly 0-4°C) and only enables brief periods of preservation (4-6 hours for the heart and 12-18 hours for the liver[7]) before the organ expires and is unfit for transplantation. Due to the Arrhenius-like nature of metabolic reactions, reducing the preservation temperature further could extend these durations, and as result, remove significant logistical hurdles, improve immunological matching and expand access to life-saving organs[8]. Critically however, the freezing point of water (0°C) represents a fundamental barrier, as freezing of biological systems is often fatal. Reducing the temperature of preservation below 0°C without ice formation thus holds significant implications for global health.

Despite manifesting at the human and even global scale, the fundamental process of nucleation involves the coordinated dynamics of a handful of molecules occurring at or even below the nanoscale[1,9]. This immense separation of scales results in making predictions about how nucleation occurs extremely difficult. Additionally, nucleation occurs almost always heterogeneously, which means that not only the physics of the liquid itself must be understood but also the interactions with all surfaces it finds itself in contact with. The result is that, depending on the catalyzing efficiency of the nucleating surfaces present in a given volume of water, ice may nucleate anywhere from just below the equilibrium melting point of 0°C down to the homogeneous nucleation limit around –40°C (pure water at atmospheric pressure)[9]. Likewise, when held at a constant subcooled temperature, water may freeze within seconds or remain in a metastable liquid state for years at a time.

In theory, if one were to characterize every surface present in a system, then the nucleation behavior could be predicted for that system and subsequent systems of the same heterogeneous composition. This has been the approach of many researchers, for example, in the atmospheric sciences, who have thoroughly quantified the nucleation behavior of many common aerosol particulate materials in order to understand larger phenomena such as cloud formation[10]. These studies often seek to describe the time dependent nature and inter-sample variability observed in freezing experiments, capturing active site variability through distributions of contact angle, surface area, or combinations thereof[11–14].

This scenario is uncommon however, since the typical engineer seeking to predict the freezing behavior of arbitrary systems cannot realistically characterize each and every system, and especially not *in situ*[15]. What's more, nucleation is known to be sensitive to variation in impurities to such an extent that the nucleation behavior of samples taken from a single source may even vary significantly from one another[16]. This phenomenon is known as non-self-averaging behavior and results from catalyst sites possessing a wide distribution of nucleation rates[16–18]. Since the nucleation rate is proportional to the exponential of the nucleation barrier, large variations in characteristic nucleation efficiency between active sites can result in one or a few active sites making up the entirety of the total system nucleation rate. This is thought to be responsible for the common experimental observation of nucleation occurring repeatably at individual sites on a surface[19,20].

The recognition that the single most potent active site may be responsible for nucleation is equivalent to the so-called weakest-link hypothesis common in the study of fracture mechanics[21] (i.e., a chain is only as strong as its weakest link, and similarly, a material placed under stress will fracture due to its weakest defect). The weakest link analogy is an example of the statistical theory of extreme values[22], which describes situations with a large number of random variables, $N$, for which we are interested in the probability distribution of the largest (or smallest) of the $N$ random variables. Extreme value statistics is also used to model, for example, extreme weather and financial events[22].

In the study of nucleation, the smallest nucleation barrier, highest nucleation temperature, or shortest nucleation induction time are relevant extreme variables. Extreme value theory can thus be recognized as the foundation of the classic singular model for heterogeneous nucleation which describes nucleation by a single characteristic nucleation temperature[23,24]. The singular model has been found to describe the distribution of nucleation temperature among a collection of transiently cooled droplets and to predict how nucleation temperature changes with the size of droplets[25,26].

Although this formulation of the classic singular model does not describe a time dependence of nucleation, extreme value theory has also been applied to experiments that measure the time until nucleation occurs for droplets held at a constant temperature. For systems exhibiting non-self-averaging behavior, the measured nucleation rate of a collection of droplets strays from the exponential decay predicted by Poisson statistics, instead becoming a stretched exponential[27].

Ultimately, the randomness of nucleation represents a far-reaching scientific challenge and specifically poses significant barriers to the deployability of supercooling for the preservation of biological systems in clinical settings. While the inherent stochastic nature of nucleation results in a finite probability for nucleation at all times when in a supercooled metastable state, at low undercoolings, and when adequately characterized, the probability is vanishingly small[28]. On the other hand, the heterogeneous catalysis of nucleation by random active sites adds an additional layer of unpredictability to supercooling stability. In even the purest volumes of water, the presence of minute insoluble impurities as well as certain soluble macromolecules can result in widely varying freezing behavior even in samples from the same source.

In this study, we develop a statistical model of nucleation rooted in both the observed inter-sample variability and the intra-sample stochasticity. The time and temperature dependent stochasticity is captured by conventional Poisson statistics, and the active site variability is captured by a new bivariate extreme value model describing variability of both the kinetic and thermodynamic heterogeneous nucleation parameters. By characterizing both sources of randomness, this study uncovers fundamental properties of heterogeneous nucleation and provides a basis for the rational design of safe and robust supercooled biopreservation protocols.

## II. METHODS

### A. Supercooling experiments

The principal goal of supercooled biopreservation is to stably hold biological material in a supercooled liquid state in order to suppress metabolic activity and achieve a certain degree of suspended animation. Nucleation is a random process and directly characterizing supercooling stability would require conducting experiments under the experimentally relevant timescales (days, possibly weeks) and repeating experiments many times in order to build statistical confidence. Because this is generally impractical, we may alternatively seek to indirectly predict supercooling stability by characterizing the system nucleation rate. The nucleation rate describes the rate of formation of critical ice clusters in a supercooled liquid and is related to nucleation probability under the framework of Poisson statistics (see Section II.B).

Experiments for characterizing nucleation rates generally fall under two categories: 1) isothermal and 2) constant cooling rate. In isothermal experiments, a sample is quickly quenched to and held at a constant supercooled temperature, and the time at which nucleation occurs is recorded. In constant cooling rate experiments, the temperature of a sample is continually decreased at a constant rate, and the temperature at which nucleation occurs is recorded. Through the application of Poisson statistics, equivalent information (namely the nucleation rate) can be obtained from these two techniques. In this study, constant cooling rate experiments were chosen for collecting the bulk of nucleation data since they allow data to be collected more quickly. Since isothermal supercooling experiments replicate the scenario employed during supercooled biopreservation, an additional campaign of isothermal experiments was performed at various temperatures and compared to the statistical predictions of the model formed from the constant cooling rate data.

The experimental setup employed in this study has been previously described by Consiglio et al.[29] and consists of a rigid Al-7075 chamber with petrolatum coated surfaces (Vaseline, Unilever, UK). The chamber is filled completely with liquid, excluding all bulk gas. The combination of hydrophobic surfaces and rigid confinement has been found to enhance the stability of supercooled solutions[29–32] and is employed here since the practical objective of this work is designing robust supercooled biopreservation protocols[28,33]. We also hypothesize that these conditions suppress nucleation from the container surface and from external perturbations and thus isolate the effect of nucleation catalysts present in the solution. The petrolatum coating, which is freshly applied before each experiment, has the additional benefit of providing maximal consistency across repeated experiments and devices.

The primary set of experiments, from which the statistical models are developed, were conducted in a 5ml chamber ($V$=5.3ml, $SA$=19cm$^2$) with a cooling rate of 2°C/min. A second set of experiments were conducted in a 20ml chamber ($V$=12.2ml, $SA$=41.9cm$^2$) with a cooling rate of 1°C/min. Within each constant cooling rate experiment, an average of 80 freezing temperatures were recorded. In order to study the variability of heterogeneous catalysts present across the samples, these experiments were repeated a total of 23 times in the 5ml systems and 10 times in the 20ml systems. Isothermal experiments were conducted in the 5ml systems at temperatures of –12°C, –13°C, and –14°C. A cutoff of two hours was imposed in order to maintain experimental throughput. All experiments were conducted with phosphate-buffered saline (PBS (1X) ID: 10010023, Thermo Fisher Scientific, USA), the melting point of which is approximately –0.5°C.

**B. Poisson statistics of nucleation processes**

Nucleation, considered here as the steady state formation of stable solid clusters, is a stochastic process with individual nucleation events being independent of each other and of previous events. From a probabilistic standpoint, it can be shown that nucleation in an ensemble of supercooled molecules is governed by Poisson statistics [34]. This yields the result that, for a volume of uniform temperature and composition, the probability of observing $k$ nucleation events in the time interval $[0, t]$ is given by the Poisson distribution:

$$P_k(t) = \frac{(\lambda t)^k}{k!} e^{-\lambda t} \qquad 1$$

where $\lambda$ is a rate parameter $[s^{-1}]$ and $t$ is time $[s]$. Since the rate parameter is constant here, Equation 1 describes a homogeneous Poisson process (note that this is distinct from the homogeneous descriptor of nucleation occurring in the bulk of a fluid and not at an interface).

From this, we can find that the probability of zero nucleation events ($k = 0$) occurring in a system with constant nucleation rate $J$ [s$^{-1}$] follows an exponential decay:

$$P(t) = \exp(-Jt) \qquad 2$$

This probability distribution is shown in Figure 1a as a function of time for three different nucleation rates. If the nucleation rate is not constant but instead varies with time, we have a nonhomogeneous Poisson process, and the probability of zero nucleation events becomes

$$P(t) = \exp\left(-\int_0^t J(\tau)d\tau\right) \qquad 3$$

In certain scenarios, the temperature is lowered at a constant cooling rate, $\dot{T} = |dT/dt|$, and we can define the probability as a function of temperature instead of time by integrating the nucleation rate in temperature, beginning at the equilibrium melting temperature, $T_m$.

$$P(T) = \exp\left(-\frac{1}{\dot{T}}\int_{T_m}^{T} J(T')dT'\right) \qquad 4$$

In statistical terms, this probability distribution is known as the survival function, $S$, which is defined as $S_X(x) = \Pr(X \geq x)$ and represents the probability that a random variable, $X$, will take on a value greater than or equal to $x$. In the context of nucleation, the survival function represents the probability that nucleation does not occur before a given time (Equation 2 for constant temperature) or degree of supercooling (Equation 4 for constant cooling rate). The survival function is related to the cumulative distribution function (CDF), $F_X(x)$, by $F_X(x) = 1 - S_X(x)$.

The probability density function (PDF) is another useful statistical function and is defined as $f_X(x) = \frac{d}{dx}F_X(x)$. Since the probability for a continuous random variable to take on any one particular value is zero, in the context of nucleation, the PDF represents the relative likelihood that nucleation will occur around a given temperature (for constant cooling rate) or around a certain time (at constant temperature). For a constant cooling rate process, the PDF is given by

$$f(T) = \frac{1}{\dot{T}}J(T)\exp\left(-\frac{1}{\dot{T}}\int_{T_m}^{T} J(T')dT'\right) \qquad 5$$

In order to solve for the freezing probability distribution, a relation for the nucleation rate $J(T)$ is needed. Classical nucleation theory (CNT) provides a framework for describing the nucleation process[35] and can be used to describe the heterogeneous nucleation rate [m$^{-2}$ s$^{-1}$] as[36,37]

$$J(T) = n_s \frac{k_B T}{h}\exp\left(-\frac{\Delta F_{\text{diff}}}{k_B T}\right)\exp\left(-\frac{\Delta G^*_{\text{hom}}f_{\text{het}}}{k_B T}\right) \qquad 6$$

where $n_s$ is the number of water molecules per unit area in contact with the ice nucleus, $k_B$ is Boltzmann's constant, $h$ is Planck's constant, $\Delta F_{\text{diff}}$ is the energy of activation for a water molecule to cross the water/ice interface, $\Delta G^*_{\text{hom}}$ is the homogeneous free energy of formation for a critical ice embryo, and $f_{\text{het}}$ is a geometric compatibility factor describing the reduction of the free energy barrier due to the presence of an ice nucleating surface.

Many of the terms in Equation 6 are independent of or only weakly dependent on temperature, and the nucleation free energy barrier, $\Delta G_{\text{hom}}$, often dominates the variation of the nucleation rate with temperature[38]. By grouping together the kinetic terms and approximating the temperature dependence of the free energy barrier as $\Delta G_{\text{hom}} \sim (T \Delta T)^{-2}$, we can arrive at an oft employed two-parameter relation[39,40]:

$$J(T) = A \exp\left(\frac{-B}{T^3 \Delta T^2}\right) \qquad 7$$

where $\Delta T = T_m - T$ is the degree of supercooling, and $A$ and $B$ are constants. The parameter $A$ is often referred to as the pre-exponential factor or kinetic term [s$^{-1}$], and the parameter $B$ as the nucleation barrier or thermodynamic term [K$^5$]. With this relation, we arrive at cumulative distribution and probability density functions, respectively, for a constant cooling rate process:

$$F(T|A,B) = \exp\left[-\frac{A}{\dot T}\int_{T_m}^{T} \exp\left(\frac{-B}{T'^3 \Delta T'^2}\right) dT'\right] \qquad 8$$

$$f(T|A,B) = \frac{A}{\dot T}\exp\left(\frac{-B}{T^3 \Delta T^2}\right)\exp\left[-\frac{A}{\dot T}\int_{T_m}^{T} \exp\left(\frac{-B}{T'^3 \Delta T'^2}\right) dT'\right] \qquad 9$$

The cumulative distribution and probability density functions for a characteristic system with $T_m = 0°C$ and nucleation parameters $A = 1.5 \times 10^{19} \text{s}^{-1}$ and $B = 1.5 \times 10^{11} \text{K}^5$ are shown in Figure 1b. The average nucleation temperature and standard deviation for this process are –13.7°C and 0.2°C, respectively.

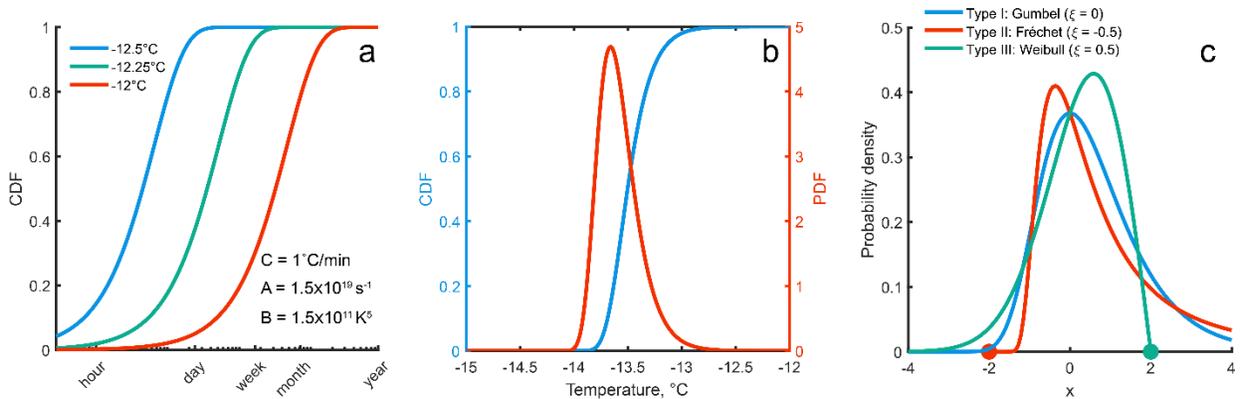

**Figure 1: Probability distribution functions.** a) Isothermal cumulative distribution functions for a system with equilibrium melting temperature $T_m = 0°C$ and nucleation parameters $A = 1 \times 10^{19} \text{ s}^{-1}$ and $B = 1.5 \times 10^{11} \text{ K}^5$ at temperatures –12.5°C (blue), –12.25°C (green), and –

12°C (red), computed from Equations 2 and 7. b) Cumulative distribution (blue) and probability density (red) functions for same system as in panel a) cooled at a constant rate of 1°C/min, computed from Equations 8 and 9, respectively. c) Example Type I: Gumbel (blue), Type II: Fréchet (red), and Type III: Weibull (green) generalized extreme value probability density functions (Equation 13) for $\boldsymbol{\mu = 0}$ and $\boldsymbol{\sigma = 1}$.

The moments of a probability distribution provide useful quantitative measures about the shape of the distribution. The first moment describes the expected value (i.e., mean), the second moment is related to the variance, and the third moment is related to the skewness. The $n$-th central moment is described generally by

$$\mu_n = \int_{-\infty}^{\infty} x^n dF(x) = \int_{-\infty}^{\infty} x^n f(x) dx \qquad 10$$

The goal of nucleation experiments, to quantify the nucleation rate, is achieved by fitting the cumulative distribution function from Equation 8 to an empirical distribution of nucleation temperatures from a constant cooling rate experiment. This nucleation rate can then be used with Equation 3 to predict the probability that this system would freeze for an arbitrary thermal history as a function of time.

Although this procedure can be used to accurately characterize individual samples, its extension to the prediction of freezing probabilities for arbitrary systems (of nominally equivalent composition) is complicated by the fact that nucleation in real bulk systems is a heterogeneously catalyzed process. Samples of water, even when extensively purified[41], inevitably contain varying amounts and populations of heterogeneous nucleating agents, and this often results in different nucleation rates between seemingly identical samples[16,18].

In order to characterize the freezing probability for heterogeneous nucleation, one would ideally know properties of every nucleating agent present. The total system nucleation rate could then be expressed as the sum of the individual rates:

$$J(T) = \sum_i^N A_i \exp\left(\frac{-B_i}{T^3 \Delta T^2}\right) \qquad 11$$

where $N$ is the total number of active sites and $A_i$ and $B_i$ are the nucleation parameters of the $i$-th active site. Active sites are in some instances considered zero dimensional entities, such as the case for a wedge or crevice, and in this case, their surface area is not a relevant parameter. Otherwise, the term $A_i$ would contain the total surface area of the active site. Equation 11 can be used in place of the single nucleation rate from Equation 7 in the above statistical relations.

This approach has been widely adopted by the atmospheric science community to characterize the nucleating behavior of specific aerosol materials[37]. In these experiments, precise quantities of rigorously characterized material are added to the system. The nucleation rate of this system can then be measured, and any observed difference can be attributed to the added material if the nucleating agents result in nucleation at higher temperatures than the base solution[42–44].

Unfortunately, this methodology cannot be applied if the purity of a solution or exact properties of all active heterogeneous catalysts are not known (such as the case with a bottle of standard laboratory-grade phosphate-buffered saline). Additionally, if the nucleation rate is strongly dependent on surface features as well as surface chemistry, the observed nucleation rate between samples of the same absolute composition may not be consistent[20,45]. Very few tools are at our disposal for predicting nucleation in these instances. In the following section, we leverage the observation that nucleation often occurs at the single most potent active site[18,24,46], enabling the application of methods from extreme value statistics for describing the effect of random impurities and circumventing the need of fully characterizing every impurity.

C. **Extreme value statistics of heterogeneous nucleation**

Any model of heterogeneous nucleation must describe, in some manner, the inherent variability of nucleation active sites that contributes to the total system nucleation rate as described in Equation 11[47]. When specified quantities of characterized impurities are added to a solution, the nucleation parameters are known (i.e., can be characterized). Conversely, in the case of a system containing uncharacterized impurities, the total number of active sites, $N$, as well as the nucleation parameters, $A$ and $B$, become random variables[18].

A common experimental observation is that ice nucleates repeatably at individual sites, such as a scratch on a container's surface[19,41]. As discussed earlier, this behavior is characteristic of quenched disorder and often leads to non-self-averaging behavior. The result is a nucleation behavior varying randomly between seemingly identical samples.

The statistical theory of extreme values has been proposed to describe this nucleation phenomenon and is at the core of first statistical theory of heterogeneous nucleation[23,24,26]. The theory is described generally as follows: a series of independent and identically distributed random variables, $X_1, \ldots, X_n$, is described by the cumulative distribution function, $F_X(x)$, and maximum, $M_n = \max(X_1, \ldots, X_n)$. Although the distribution of the maximum value of this set is exactly given by $\Pr(M_n \leq z) = [F(z)]^n$, the distribution $F_X(x)$ is often not known such as in the case of heterogeneous nucleation with a distribution of active sites. In the asymptotic limit of $n \to \infty$ however, the distribution of the maximum is described, regardless of the underlying distribution $F_X(x)$, by a finite family of distributions known as the generalized extreme value (GEV) distribution[22,48]:

$$G(x) = \exp\left\{-\left[1 + \xi\left(\frac{x-\mu}{\sigma}\right)\right]^{-\frac{1}{\xi}}\right\} \qquad 12$$

$$g(x) = \frac{1}{\sigma}\left[1 + \xi\left(\frac{x-\mu}{\sigma}\right)\right]^{-\left(1+\frac{1}{\xi}\right)} \exp\left\{-\left[1 + \xi\left(\frac{x-\mu}{\sigma}\right)\right]^{-\frac{1}{\xi}}\right\} \qquad 13$$

Here, $G(x)$ is the cumulative distribution function, $g(x)$ is the probability density function, and $\mu$, $\sigma$, and $\xi$ are the location, scale, and shape parameters, respectively. The specific distribution is determined by the shape parameter, which governs the tail behavior. While the Gumbel distribution is unbounded in both limits, the Weibull distribution has an upper bound and the Fréchet distribution a lower bound. The sub-classes of the GEV distribution are summarized in Table 1 and depicted in Figure 1c for location and scale parameters $\mu = 0$ and $\sigma = 1$, respectively. Even though we cannot realistically know the complete distribution of impurities present in a solution, in the asymptotic limit, the distribution of most potent sites is predicted to follow one of these three distributions.

In order for extreme value theory to be applied to heterogeneous nucleation experiments, a few additional assumptions must first be introduced: 1) the variability in the number of active sites between systems is much smaller than the total number of active sites, and 2) the underlying distribution of active sites is the same between systems. These assumptions are reasonable for individual samples taken from the same source.

**Table 1: Generalized extreme value distribution sub-types.**

| Type | Name | Shape factor | Cumulative distribution function | |
|---|---|---|---|---|
| I | Gumbel | $\xi = 0$ | $G(x) = \exp\left[-\exp\left(-\frac{x-\mu}{\sigma}\right)\right]$ | 14 |
| II | Fréchet | $\xi > 0$ | $G(x) = \begin{cases} \exp\left[-\left[1+\xi\left(\frac{x-\mu}{\sigma}\right)\right]^{-1/\xi}\right] & x > 0 \\ 0 & x \leq 0 \end{cases}$ | 15 |
| III | Weibull | $\xi < 0$ | $G(x) = \begin{cases} \exp\left[-\left[1+\xi\left(\frac{x-\mu}{\sigma}\right)\right]^{-1/\xi}\right] & x < 0 \\ 1 & x \geq 0 \end{cases}$ | 16 |

1. **Determining the relevant extreme variables**

The specific physical quantity that is chosen as the extreme variable confers additional constraint on the problem. For example, the classic singular model of nucleation prescribes a characteristic activation temperature to each active site, and thus the nucleation temperature is the extreme variable. In reality, nucleation is stochastic and does not occur at a single temperature but rather over a distribution of temperatures, even for a single active site. For extreme value statistics to be a good descriptor of the nucleation temperature, the variability of nucleation temperature for individual active sites should be small in comparison to the variability of characteristic nucleation temperature between systems. This is the case for the system studied here, as seen in Figure 2 and Figure 3, where the standard deviation of nucleation temperatures for a single sample is on average about 0.33°C while the standard deviation of nucleation temperatures between samples is approximately 1.5°C.

The classic singular model is typically applied to data from constant cooling rate experiments in which nucleation temperature is the measured quantity. Alternatively, in the case of isothermal experiments, the relevant extreme variable is the nucleation induction time, and the active site with the minimum characteristic induction time is responsible for nucleation[49]. For extreme value statistics to have reasonable predictive power here, the spread of induction times for individual active sites caused by inherent stochasticity should likewise be small compared to the spread of induction times between samples.

These two applications of extreme value statistics are the most readily applied since the extreme variable is a measured quantity. Alternatively, we can consider the molecular mechanisms of nucleation and instead of attributing nucleation to the active site with the highest nucleation temperature or shortest nucleation induction time, we can describe the potency of the active site by the nucleation rate itself. In this way, nucleation is attributed to the active site with the largest nucleation rate. In the two-parameter nucleation rate relation described in Equation 7, the nucleation barrier, $B$, is exponentiated, and as such, the nucleation rate is much more sensitive to $B$ than it is to the pre-exponential factor, $A$. A theoretical extreme value analysis of nucleation performed by Sear[16,18] held the pre-exponential factor constant and treated the nucleation barrier as the extreme variable. This approach is convenient since it reduces the problem to a single variable, however it does not agree well with the common experimental observation that the spread of nucleation temperatures is rather constant regardless of the temperature at which nucleation occurrs[50]. Both the expected value and standard deviation of the nucleation temperature for a constant cooling rate experiment, as computed from the first and second moments of the probability distributions, are dependent on pre-exponential and nucleation barrier parameters. Therefore, both nucleation parameters must vary for the standard deviation to remain the same as the nucleation temperature changes.

Since none of the singular extreme value models can suitably capture the time and temperature dependence of nucleation nor the variability of the nucleation rate, we propose a hybrid scheme that treats both the nucleation temperature and the nucleation barrier parameter, $B$, as extreme quantities. This bivariate approach is then able to account for the variability of the nucleation parameter, $A$, by leveraging the definition of the average nucleation temperature obtained as the first central moment of the constant cooling rate probability distribution function.

$$\langle T_f \rangle = \int_{-\infty}^{\infty} T f(T) dT = \int_{-\infty}^{\infty} \frac{A}{\dot{T}} \exp\left(\frac{-B}{T^3 \Delta T^2}\right) \exp\left[-\frac{1}{\dot{T}} \int_{T_m}^{T} A \exp\left(\frac{-B}{T'^3 \Delta T'^2}\right) dT'\right] T \, dT \quad 17$$

An additional limitation of the classic singular model, which prescribes a single activation temperature to nucleation sites, is that it does not capture the cooling rate dependence of the nucleation process. This limitation is overcome in this hybrid model since the expression relating the extreme variables, $\langle T_f \rangle$ and $B$, to the nucleation rate pre-exponential factor, $A$, in Equation 17 incorporates the cooling rate.

The proposed hybrid scheme is a case of bivariate analysis, with each extreme variable described by separate univariate distributions. The nucleation temperature and nucleation barrier are not completely independent variables however, so the problems cannot be treated separately. Instead, we must define a joint probability distribution, considering the nucleation temperature and nucleation barrier as independent yet correlated variables. Formulation of bivariate extreme value distributions with arbitrary marginal distributions is in general difficult due to the construction of the dependence structure and the absence of a finite parametric family, which exists for univariate extreme value distributions. The two most common bivariate models are the mixed and logistic models. Here, we implement the logistic model[51–53]:

$$G(x,y) = \exp\left\{-\left([-\ln(F_1(x))]^m + [-\ln(F_2(y))]^m\right)^{\frac{1}{m}}\right\} \quad 18$$

where $G(x,y)$ is the bivariate extreme value distribution describing both random extreme variables, $x$ and $y$, and $F_1(x)$ and $F_2(y)$ are the univariate marginal extreme value distributions. The quantity $m$ is a measure of the dependence between the random variables and is defined as

$$m = \frac{1}{\sqrt{1-\rho}} \qquad 19$$

where $\rho$ is the correlation coefficient. A value of $m = 1$ corresponds to independence of the two variables and leads to the joint distribution $G(x,y) = F_1(x)F_2(y)$. The joint probability density function follows similarly from the univariate definition and is given by

$$g(x,y) = \frac{\partial^2 G}{\partial x \partial y} \qquad 20$$

2. **Determining the relevant extreme value distribution**

Now that we have identified the extreme parameters and determined the form of the bivariate distribution, the next task is to identify the type of GEV distribution that describes each of the marginal distributions. Although the asymptotic theory predicts that the process will converge to one of the three types of GEV distributions shown in Table 1, we cannot immediately determine which one this is without information about the underlying process or distribution. If the underlying distribution of active sites is known, then the distribution of maximums is also known, and an extreme value analysis is moot. In the worst-case scenario, we cannot make any physical suppositions and are therefore left to fit data directly to the full three-parameter GEV distribution. By considering some properties of the GEV sub-classes however, we may be able to rule out some cases. For instance, we can consider the tail behavior of the distributions.

Both the nucleation temperature and the nucleation barrier are physically bounded. The nucleation temperature cannot be higher than the equilibrium melting point and cannot be lower than the homogeneous limit (roughly –40°C). In bulk systems that are not purified extensively, heterogeneous nucleation usually results in freezing occurring closer to the equilibrium melting point, making the Weibull distribution potentially a natural fit. Similarly, the nucleation barrier must be less than the homogeneous limit but must also be greater than zero. Since the nucleation barrier often varies many orders of magnitude in value, it is convenient to consider the logarithm of the value (i.e., $\ln B$). The logarithm of the nucleation barrier does not inherently have a lower bound (since extreme value analysis deals with maximums, we must consider $-\ln B$ which conversely does not have an inherent upper bound). We can most likely rule out the Weibull distribution, leaving the Fréchet distribution as a plausible choice for the log-distributed nucleation barrier. In this case, the homogeneous nucleation barrier would provide a physically sensible lower limit.

Often the best option for assessing which probability distribution best describes the distribution is using the graphical method of the probability plot[22,53]. A good fit is indicated by the formation of a straight line when plotting the quantiles of the empirical distribution against the quantiles of a reference distribution. The probability plots in Figure 3 are linear in nature, indicating a decent goodness of fit has been obtained for the measured nucleation parameters. Since there is uncertainty in the underlying active site distribution and little data in the literature on extreme value distributions for either nucleation temperatures or nucleation rate parameters, it is still recommended that each extreme value distribution be explored when studying a new system.

## III. RESULTS & DISCUSSION

### A. Results from constant cooling rate experiments

The survival curves of the nucleation temperatures are shown in Figure 2 for a series of 23 constant cooling rate experiments performed in the 5ml system with PBS at 2°C/min. While the spread of nucleation temperatures for individual trials is typically within 1°C (average standard deviation of 0.33°C), nucleation temperatures across the set of experiments ranges nearly 7°C (standard deviation of average nucleation of 1.55°C). This is consistent with nucleation experiments in the literature that find narrow spreads for individual systems and large spreads for experiments conducted on multiple systems[54,55]. The stochastic behavior for individual nucleating catalysts is thus quite narrow, yet the characteristic temperature around which this purely stochastic behavior is centered can vary multiple degrees from system to system: a hallmark of non-self-averaging behavior, which indicates the potential suitability of extreme value theory.

The method of maximum likelihood estimation is used to determine the cumulative distribution function from Equation 7 and obtain the respective nucleation parameters for each of the survival curves. As a check of whether the observed nucleation process is well described by the proposed empirical nucleation rate $J(T) = A \exp(-B/T^3 \Delta T^2)$, we can linearize the constant cooling rate CDF (see Deubener and Schmelzer[56]). The linearized nucleation spectra are shown in Figure 1b, and their linear form suggests the observed nucleation process is generally well described by the proposed nucleation rate.

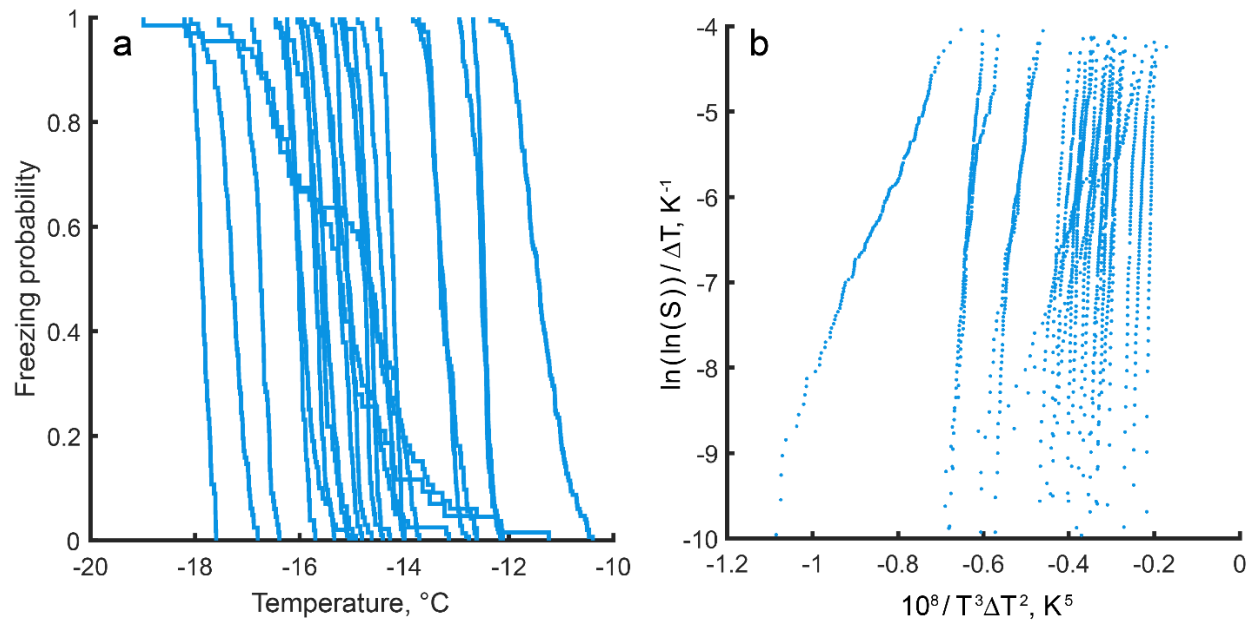

**Figure 2: Constant cooling rate nucleation temperature data.** a) Empirical cumulative distribution functions for a series of 23 constant cooling rate experiments performed in the 5ml system with PBS at 2°C/min. b) Survival curves from panel a) linearized with respect to the cumulative distribution for the constant cooling rate experiment described by the nonhomogeneous Poisson process in Equation 7.

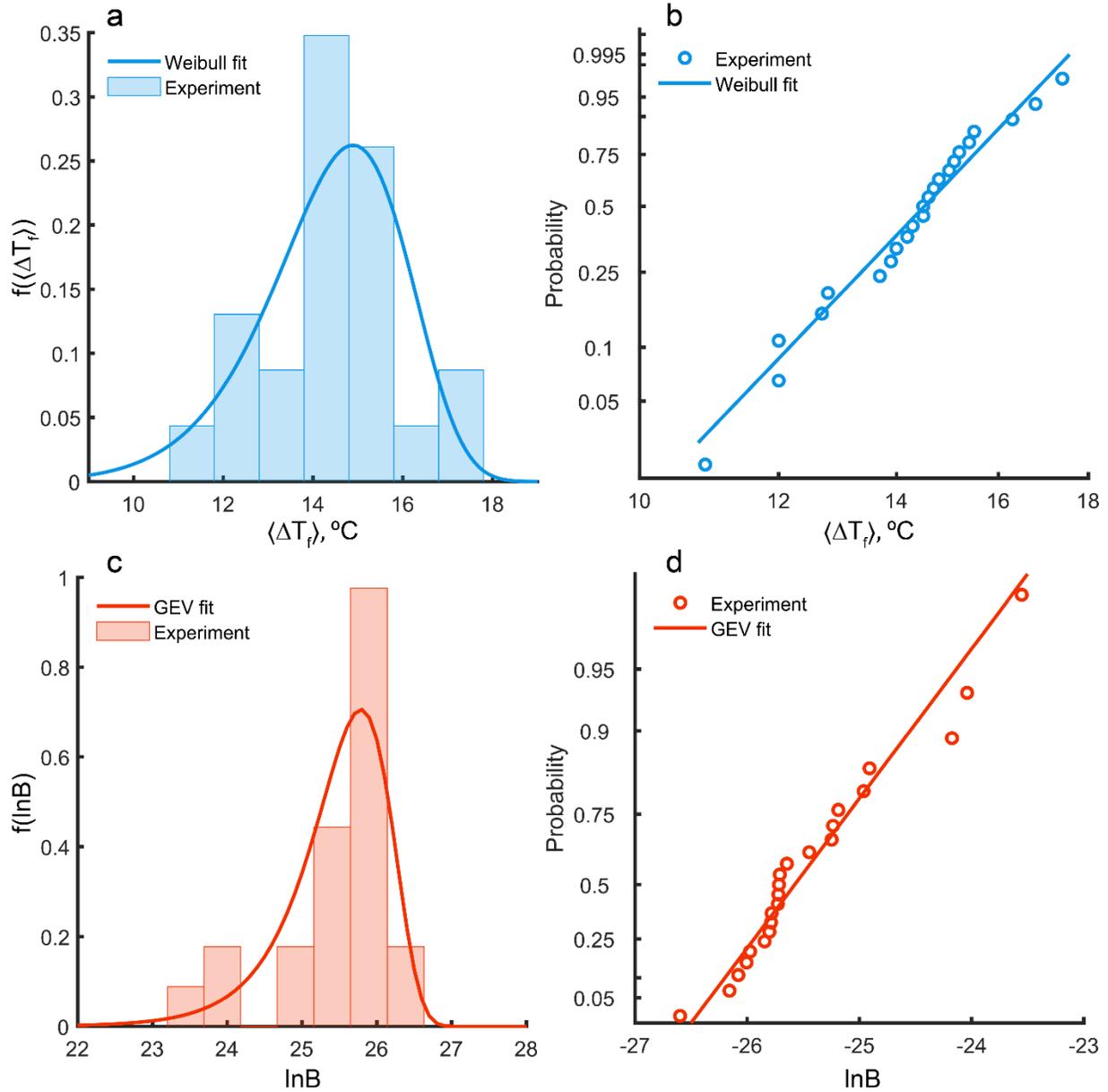

**Figure 3: Extreme value distributions of mean nucleation temperature and nucleation barrier.** a) Histogram and Weibull PDF of average nucleation temperatures from 5ml constant cooling rate experiments (n=23). b) Weibull probability plot of average nucleation temperatures. c) Histogram and GEV PDF for nucleation barrier from same experiments. d) Fréchet probability plot of nucleation barrier parameter.

**B. Distribution of extreme parameters**

With the empirical survival curves and inferred nucleation parameters in hand, we can now evaluate the distributions of the extreme variables: average nucleation temperature, $\langle \Delta T_f \rangle$, and

13 | 24

nucleation barrier, $\ln B$. Maximum likelihood estimation is used again for fitting of the extreme value distributions and the resulting distributions are shown in Figure 3 for the 5ml constant cooling rate experiments.

As discussed in Section II.C.2, the Weibull distribution is a physically plausible choice for describing the distribution of nucleation temperatures between samples because of the melting temperature upper bound. A slight modification is first made to the Weibull distribution as defined in Equation 16. By considering the degree of supercooling upon freezing, $\Delta T_f = T_m - T_f$, instead of the absolute nucleation temeprature, the location of the upper bound is shifted from $T_m$ to zero. This effectively reduces the number of parameters in the Weibull distribution from three to two. Figure 3a depicts a histogram of the average nucleation temperatures along with the probability density function of the Weibull fit, and Figure 3b shows the Weibull probability plot of the average nucleation temperatures. Linearity in the probability plot indicates the $\langle \Delta T_f \rangle$ data are well described by the Weibull distribution. We find an average degree of supercooling upon freezing of $\langle \Delta T_f \rangle = 14.3°C$ and standard deviation of $\sigma_{\langle \Delta T_f \rangle} = 1.6°C$.

In the case of the nucleation barrier, the full three parameter GEV distribution is fit to the nucleation barrier data and a best fit is found with the Fréchet distribution. This agrees with the intuition discussed in Section II.C.2 that the homogeneous nucleation barrier represents a natural limit. Wood and Walton[39] found a value of $B_{hom} = 1.66 \times 10^{12} K^5$ ($\ln B_{hom} = 28.14$), and though this limit has not been strictly enforced when fitting the Fréchet distribution here, the probability density function evaluated at the homogeneous limit is on the order of $10^{-60}$ and is small enough to exclude any nonphysical contributions. Figure 3c depicts a histogram of the nucleation barriers along with the probability density function of the fitted Fréchet distribution. Figure 3d shows the Frechet probability plot and the observed linearity suggests an appropriate fit. We find an average nucleation barrier of $\langle \ln B \rangle = 25.4$ and standard deviation of $\sigma_{\ln B} = 0.7$.

Having determined the marginal distributions, we can now consider the joint distribution. To that end, we first compute the correlation parameter for the two extreme variables using Equation 19 and find a value of $m = 1.32$. The joint probability density function, $g(\langle \Delta T_f \rangle, \ln B)$, is then calculated from Equations 18 and 20 and is shown in Figure 4a. Next, we can perform a variable substitution in order to obtain the PDF in terms of the nucleation rate parameters, $A$ and $B$, using the definition of the mean nucleation temperature form Equation 17. From the result, shown in Figure 4b, we see that the there is a narrow band of probable pairings, which indicates that nucleation parameters have a somewhat high degree of correlation and perhaps represents a fundamental property of heterogeneous nucleation.

In the context of classical nucleation theory (and shown in Equation 6), the heterogeneous nucleation barrier, $\Delta G^*_{het}$, is the product of the homogeneous nucleation barrier, $\Delta G^*_{hom}$, and a heterogeneous shape factor, $f$, which accounts for a modified contact area between the growing solid ice cluster and the catalyst particle onto which it nucleates[57]:

$$\Delta G^*_{het} = \Delta G^*_{hom} f_{het}(\theta) \qquad 21$$

The shape factor, $f_{het}$, is a function of the contact angle, $\theta$, between the ice embryo and nucleating surface and can vary between $f(0°) = 0$ (corresponding to complete wetting and no nucleation barrier) and $f(180°) = 1$ (corresponding to complete hydrophobicity and homogeneous nucleation). Hydrophilic surfaces, which have small contact angles and also smaller nucleation barriers, are often polar in nature, and polar surfaces are known to orient liquid water molecules in their close vicinity. Therefore, a potential explanation for the scaling behavior seen in Figure 5c and correlation seen in Figure 4b could be that the reduction in mobility of liquid molecules caused by more potent hydrophilic nucleating surfaces results in a smaller pre-exponential diffusive coefficient.

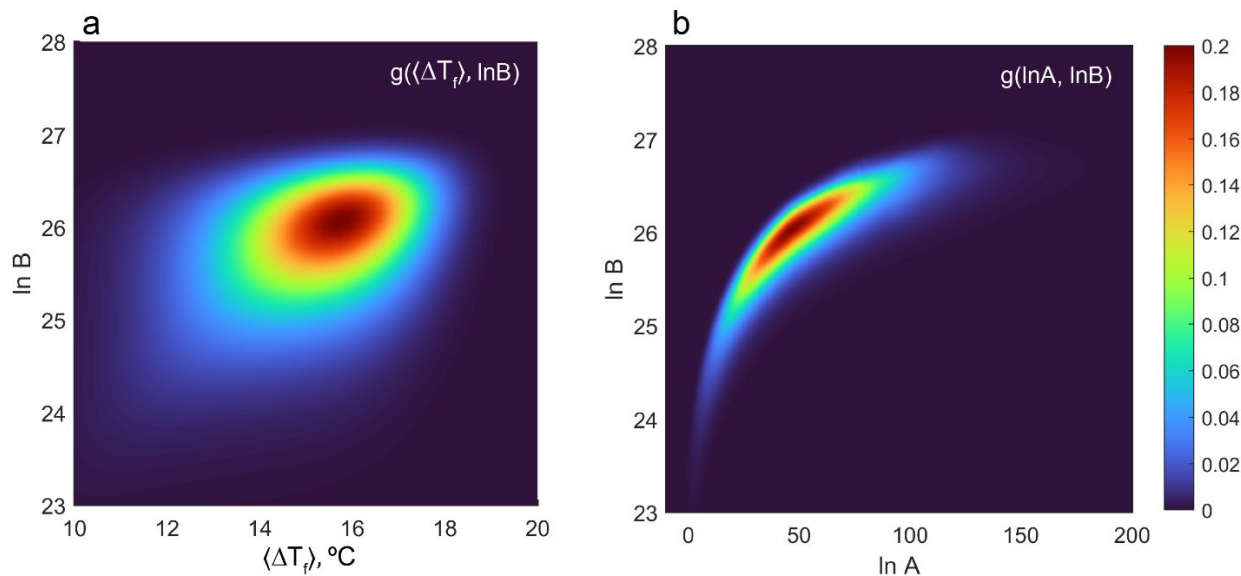

**Figure 4: Bivariate extreme value distributions.** a) Joint probability density function, $g(\langle \Delta T_f \rangle, \ln B)$ computed from the marginal univariate extreme value distributions for the mean freezing temperature, $\langle \Delta T_f \rangle$, and nucleation barrier, $B$. The distribution is centered at $\langle \Delta T_f \rangle = 14.3°C$ and $\ln B = 25.4$. b) Probability density function from panel a) with the mean freezing temperature transformed to $\ln A$ using the definition of the mean nucleation temperature for a constant cooling rate experiment from Equation 16. A narrow distribution indicates a strong correlation between the kinetic and thermodynamic nucleation parameters.

Now that we have determined the distribution of the nucleation parameters from extreme value statistics, we can apply this model to useful ends. Since the 5ml systems are not a practical size for preserving anything beyond cell suspensions and small tissue constructs, we may be interested in seeing how the model can be used to predict nucleation behavior in larger, more relevant volumes. This will be covered in the following section. Additionally, the experiments from which the model is constructed, constant cooling rate experiments, do not represent the modality of supercooling employed when actually preserving a biological specimen. In reality, preservation is conducted at a constant temperature and at temperatures warmer than the supercooling limits probed in constant cooling rate experiments. We may therefore be interested in evaluating how the model predicts nucleation behavior in isothermal systems and as a function of time. This will be covered in Section III.D.

## C. Scaling of the nucleation process with system size

A useful property of probability distributions is that the probability of independent events occurring together is the product of the independent probabilities (i.e., $P(A \text{ and } B) = P(A) \cdot P(B)$). If we consider a large system being comprised of a collection of smaller independent systems, we can generalize this rule to predict how probability distributions scale with system size. For example, if $F(x)$ is the cumulative distribution function for a system of size $n_1$, then the distribution for an identical system of size $n_2$ is $[F_1(x)]^{n_2/n_1}$. Following this logic, the size of the system can be readily incorporated into the GEV distribution yielding:

$$G(x, V) = [G(x)]^{V/V_0} = \exp\left\{-\frac{V}{V_0}\left[1 + \xi\left(\frac{x-\mu}{\sigma}\right)\right]^{-1/\xi}\right\} \qquad 22$$

where $V_0$ is the volume of the system from which the model was developed, and $V$ is the volume of an arbitrary system. If a process were to scale with surface area rather than volume, then the relevant surface area quantity could be used instead. In systems where heterogeneous catalysts are intentionally added in order to study their nucleation behavior or to initiate nucleation, the relative scaling parameter would be the total nucleant surface area which is then often converted to a concentration.

Incorporating the system size into the probability functions in this manner can help assess the validity of the statistical model developed thus far. For example, we can conduct a campaign of constant cooling rate experiments in a larger system and assess the level of agreement with the volume scaling predictions from the statistical model based on the 5ml data.

The 5ml and 20ml systems employed in this study have a volume ratio of $V_{20}:V_5 = 4$ and a surface area ratio of $S_{20}:S_5 = 2.2$. Although we can hypothesize that the petrolatum coating removes the possibility of ice nucleation on the container surface since it is a slippery hydrocarbon-based material similar to oils used to study homogeneous nucleation in emulsions, we cannot decisively conclude whether nucleation is initiated on the container surface during the constant cooling rate experiments and thus expect nucleation behavior to scale with surface area, or whether nucleation is initiated in the bulk of the fluid and thus expect nucleation to scale with volume.

An additional factor that must be considered is the temperature gradient that persists within the cylindrical volumes while being cooled at a constant rate. The 5ml chamber has an inner radius of 6.35mm and the 20ml chamber has an inner radius of 12.7mm. From a simple scaling analysis, we can approximate the temperature difference from the wall to the center axis of the cylindrical chamber, as well as the radial thermal penetration depth within which the temperature is approximately uniform. The radial thermal penetration depth for a constant wall temperature is given by $\delta_r = \sqrt{4\alpha t}$ and during constant cooling becomes:

$$\delta_r = \sqrt{\frac{4\alpha \Delta T}{C}} \qquad 23$$

where $\alpha$ is the thermal diffusivity (~0.12 mm²/s for supercooled water), $\Delta T$ is the temperature difference (°C), and $C$ is the cooling rate (°C/s). For the 5ml system and a cooling rate of 2°C/min, we find that the center axis lags the wall temperature by ~2.6°C. For the 20ml system and a cooling rate of 1°C/min, the center lags the wall temperature by ~5.2°C. In the case of volume nucleation, if we assume that nucleation is occurring in the coldest part of the system, the entire volume likely would not contribute to nucleation but rather a portion of the volume within a certain distance from the inner wall. Ultimately, the uncertainty of where nucleation occurs represents a primary limitation for analysis of nucleation in these systems and future studies should investigate this more closely.

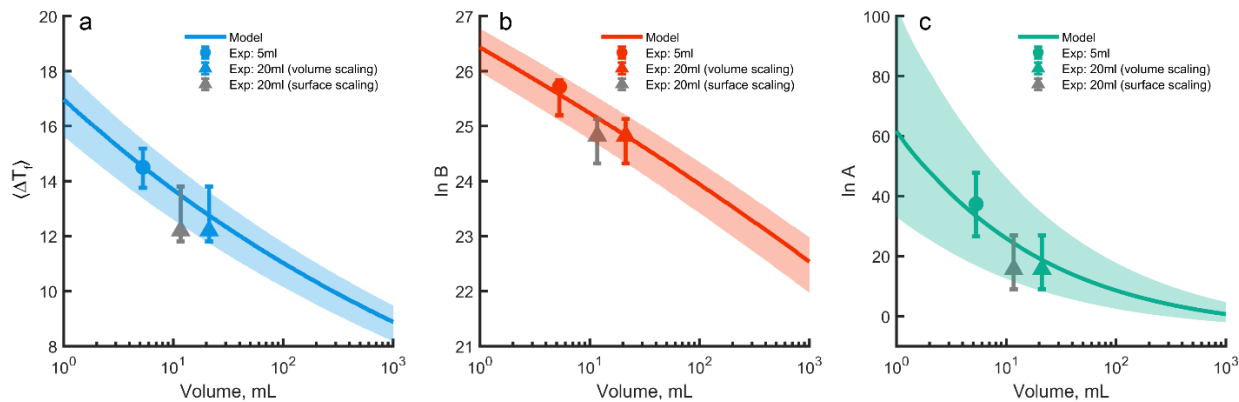

**Figure 5: Scaling of nucleation parameters with system volume.** Scaling of a) nucleation temperature, $\langle \Delta T_f \rangle$, b) nucleation barrier, $\ln B$, and c) nucleation kinetic factor, $\ln A$, computed from the individual extreme value model based on 5ml data. Kinetic parameter, $A$, is computed from the mean nucleation and nucleation barrier with the relation in Equation 16. Data points in each panel are from constant cooling rate experiments in 5ml (●) and 20ml (▲) systems. Location of the 20ml data corresponds to pure volume scaling (same color as 5ml data), and surface area scaling (gray). Solid curve denotes median value. Error bars and shaded regions denote 25%/75% quartiles.

Shown in Figure 5 are the three nucleation parameters $\langle \Delta T_f \rangle$, $\ln B$, and $\ln A$ computed as a function of volume, $V$. The solid lines represent the median value of each parameter, and the shaded regions are bounded by the 25% and 75% quartiles. The nucleation temperatures are computed by incorporating volume into the Weibull distribution fit to the 5ml data. Similarly, scaling of the nucleation barrier, $\ln B$, is computed by incorporating volume into the Fréchet distribution fit to the 5ml data. The nucleation pre-exponential factor, $\ln A$, is obtained using the relation in Equation 17.

The average nucleation temperature (Figure 5a) scales relatively linearly with the logarithm of the volume, levelling off at larger volumes. The decreasing trend of the nucleation temperature with volume is caused by an increased likelihood for more potent nucleating catalysts in larger samples. Whereas in very small systems (such as micro-emulsions whose droplet diameters are often on the order of single microns) it may be possible to exclude all or most insoluble impurities (other than the encapsulating oil phase), larger systems inherently will contain more impurities. This result is consistent with the early findings of Bigg[58,59], Langham and Mason[25], and Mossop[60] and is in accordance with the singular model for the nucleation. The linearity is expected to be a universal property for nucleation of systems containing the same population of nucleating catalysts. The precise slope and position of this curve, however, is dependent on the particular source of water and the population of impurities that it contains. The scaling of the nucleation barrier, $\ln B$, behaves similarly to the scaling of nucleation temperature, i.e., the larger the system is, the higher the likelihood of more potent nucleating agents, and thus larger systems possess smaller characteristic nucleation barriers.

One result that would be difficult to predict without this analysis is how the nucleation pre-exponential factor, $A$, scales with volume. This factor is often assumed to be the same for heterogeneous nucleation and homogeneous nucleation since in the context of CNT it is related to the diffusion of liquid molecules to the surface of the growing solid nucleus, and thus thought to be solely a property of the liquid and not of the nucleating catalyst. As discussed in Section II.C.2, for consistency of the bivariate statistical model, this parameter cannot remain constant and is actually constrained by the values of the other two parameters, $\langle \Delta T_f \rangle$ and $B$. Interestingly, the value of $A$ is predicted to decrease with increasing volume (Figure 5c), which is perhaps counterintuitive since a smaller $A$ yields a lower nucleation rate for a constant value of $B$. This behavior is in fact consistent with many studies of heterogeneous nucleation, and, as discussed in Section III.B, could be a result of reduced molecular mobility in the vicinity of potent active sites that are somewhat polar and hydrophilic. Regarding the width of the distribution, it is quite wide at small volumes and becomes quite narrow for larger volumes. It is difficult to ascribe a physical justification for this behavior since in the scope of this extreme value analysis, the parameter $A$ is not an extreme variable, and as such, we can neither predict its distribution for the most potent active site nor can we comment on the value of $A$ for the remainder of the active sites. Future studies should seek to validate this result and investigate potential physical origins.

Also shown in Figure 5 are the median parameters from experiments performed in the 5ml and 20ml systems. The 20ml data is displayed in two instances, corresponding to scaling with volume and surface area. The same general trend (increasing vs. decreasing) is seen between the experiments and model prediction which supports the underlying premises of the model. The volume scaling produces the best agreement, suggesting that nucleation is not necessarily occurring on the container's inner surface. Future experimental studies should investigate this scaling behavior further.

### D. Freezing probability versus time in systems at constant temperature

Having developed a statistical model that describes the random distribution of the kinetic and thermodynamic nucleation parameters and describes how they vary with system size, we can now address the practical objective of this research: determining the freezing probability of supercooled aqueous systems in order to identify conditions under which biological material may be held in a stable supercooled liquid state. The isothermal freezing probability, resulting from the homogeneous Poisson distribution of Equation 2 and the nucleation rate from Equation 6, is given by

$$P(T, t | A, B) = 1 - \exp\left[-A \exp\left(\frac{-B}{T^3 \Delta T^2}\right) t\right] \qquad 24$$

For systems with uncharacterized impurities, the nucleation parameters in this expression, $A$ and $B$, are random variables. The model developed in this study has treated the mean nucleation temperature and the nucleation barrier parameters as extreme variables, which enables the heterogeneous nucleation process to be described by the joint extreme value distribution in Equation 18. To incorporate the dependence on system volume, Equation 18 can be modified in accordance with Equation 22. By further transforming $\langle \Delta T_f \rangle$ to $A$ using the definition of the mean nucleation temperature from Equation 17, we can obtain an expression for the probability density function in terms of $A$ and $B$: $g(\langle T_f \rangle, B) \to g(A, B)$. Finally, we can obtain a generalized expression for the isothermal freezing probability as a function of volume and time by integrating over all possible values of $A$ and $B$ weighted by their joint probability density.

$$P(T,V,t) = \iint_{A,B} P(T,t|A,B)g(A,B|V)dAdB \qquad 25$$

This relation is used to compute the freezing probability, shown in Figure 6a, for physiological saline in a 5ml system with petrolatum-coated surfaces. We find that the logarithm of the probability scales approximately linearly with temperature for small probabilities ($p < 0.5$). The trend of freezing behavior with supercooled duration is also interesting. For a given freezing probability, varying the temperature by only a few degrees results in the corresponding induction time varying from one day to one year. This result may have the practical implication of extrapolating supercooled states observed to be stable on the order of days to being stable for much longer periods of time.

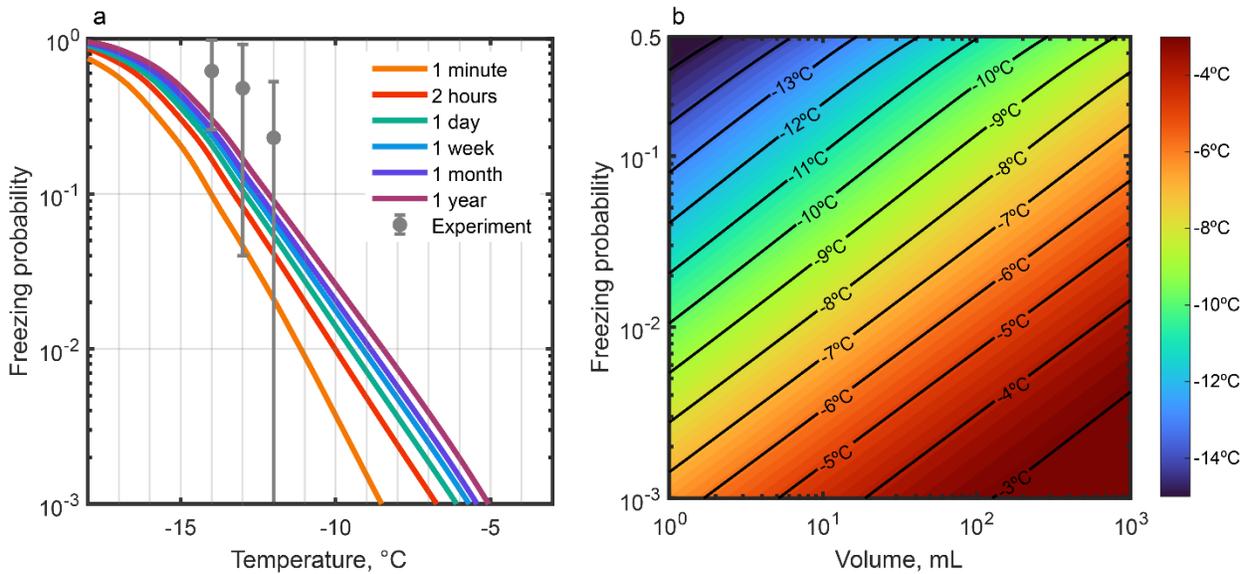

**Figure 6: Isothermal freezing probabilities.** a) Freezing probability for a rigidly confined 5ml system with physiological saline and petrolatum-coated surfaces as a function of temperature and for various durations, computed from Equation 25. Data points correspond to average freezing probabilities from the campaign of isothermal nucleation experiments with a cut-off time of two hours. Error bars correspond to +/- one standard deviation. b) Probability of freezing within a time of 24 hours as a function of system volume and temperature, computed from Equation 25.

The results of the 5ml isothermal experiments are also shown in Figure 6a. In each of these experiments, the effective quantity measured is the probability of the sample freezing within two hours of being supercooled. Average freezing probabilities of 23% (n=10), 48% (n=6), and 62% (n=18) are found for temperatures of –12°C, –13°C, and –14°C, respectively. As anticipated, the freezing probability increases with decreasing temperature, however the experimental values are consistently higher than those predicted by the model.

Considerable variation is observed for the isothermal trials (repeated experiments at individual temperatures yielded standard deviations of roughly +/–30%). Though this, as well as the relatively small number of repeated trials, likely contribute to the disagreement with predictions, this general observation is in fact in accordance with predictions of the statistical model. Due to relatively large variation in active site potency between experiments, a high likelihood is predicted of either observing almost no freezing events within 2 hours or observing a freezing rate of nearly 100%. This highlights the difficulty of performing isothermal nucleation rate experiments as compared to constant cooling rate experiments as well as the difficulty of interpreting results from individual nucleation experiments.

Figure 5b depicts the freezing probability as a function of temperature and volume for a supercooling duration of 24 hours. This heat map can be of immediate use to the applied cryobiologist by informing the choice of preservation temperature based on the size of their system and appropriate freezing risk level. For example, accepting a 1% probability compared to a 0.1% chance of freezing would enable the storage temperature to be reduced by roughly 3-4°C. Accepting a 10% chance of freezing could enable the storage temperature to be reduced an additional 3-4°C.

## IV. CONCLUSION

Supercooling, which describes the phenomenon of water remaining in a metastable liquid state below its equilibrium melting temperature, is an attractive premise for preserving biological material at low temperature yet in the absence of ice. Being a metastable process however, there is a nonzero probability at all times for freezing to occur, and this reality represents a significant challenge for the clinical translatability of any supercooling technique. Just as reactor engineers designing nuclear power plants need to know the probability of a meltdown occurring and civil engineers designing a suspension bridge need to know the likelihood of a high-magnitude earthquake, transplantation surgeons need certain assurance that their supercooled organ will not freeze during the storage period. Without methods to arrest ice growth before damage is imparted, quantifying freezing probability is expected to be central to enabling the widespread implementation of supercooling.

Ultimately, heterogeneous nucleation is a complex problem that is both incompletely understood from a molecular level and difficult to characterize from an experimental standpoint. Random active sites, existing as minute insoluble particles floating in solution or adhered to container walls as well as certain water-soluble macromolecules, can remain nearly undetectable in solution and produce seemingly unpredictable freezing behavior across repeated experiments. Predicting nucleation behavior in the presence of uncharacterized impurities remains an unsolved problem with broad implications.

In this study, we leverage the observation that nucleation often occurs repeatedly on the single most potent active site present in a system in order to approach the problem using the statistical theory of extreme values. This enables us to reduce the scope of the problem by circumventing the need to characterize every potential active site. We develop a joint singular and stochastic model based on data from constant cooling rate experiments in order to quantify the sample-to-sample variability and time-dependent intra-sample stochasticity. By capturing the variability of the kinetic and thermodynamic parameters governing heterogeneous nucleation, the model is able to predict the probability of nucleation as a function of temperature, volume, and time. Together, this approach constrains the multi-faceted probabilistic nature of heterogeneous ice nucleation and enables the rational design of supercooled biopreservation protocols.

## ACKNOWLEDGMENTS

This research received financial support from the National Science Foundation (NSF) Graduate Research Fellowship under Grant No. DGE 1752814 as well as by the NSF Engineering Research Center for Advanced Technologies for Preservation of Biological Systems (ATP-BIO) under NSF EEC Grant No. 1941543.

## AUTHOR DECLARATIONS

### Conflict of Interest

The authors have no conflicts to disclose.

### Author Contributions

AC, MPP, and BR conceived of study. AC and YO conducted experiments. AC performed statistical analyses and wrote the first draft of the manuscript. All authors contributed to revision of the manuscript.

## DATA AVAILABILITY

The data that support the findings of this study are available from the corresponding author upon reasonable request.

## REFERENCES

1. Pruppacher, H. R. & Klett, J. D. Microstructure of Atmospheric Clouds and Precipitation. 10–73 (2010). doi:10.1007/978-0-306-48100-0_2

2. Salt, R. W. Principles of Insect Cold-Hardiness. *Annu. Rev. Entomol.* **6**, 55–74 (1961).

3. Barnes, B. M. Freeze Avoidance in a Mammal: Body Temperatures Below 0°C in an Arctic Hibernator. *Science (80-. ).* **244**, 1593–1595 (1989).

4. Dalvi-Isfahan, M., Hamdami, N., Xanthakis, E. & Le-Bail, A. Review on the control of ice nucleation by ultrasound waves, electric and magnetic fields. *J. Food Eng.* **195**, 222–234 (2017).

5. Franks, F. & Jones, M. Biophysics and biochemistry at low temperatures. *FEBS Lett.* **220**, 391 (1987).

6. Giwa, S. *et al.* The promise of organ and tissue preservation to transform medicine. *Nature Biotechnology* **35**, 530–542 (2017).

7. Lewis, J. K. *et al.* The Grand Challenges of Organ Banking: Proceedings from the first global summit on complex tissue cryopreservation. *Cryobiology* **72**, 169–182 (2016).

8. de Vries, R. J. *et al.* Supercooling extends preservation time of human livers. *Nat. Biotechnol.* **37**, 1131–1136 (2019).

9. Angell, C. A. Supercooled Water. *Water Aqueous Solut. Subzero Temp.* 1–81 (1982). doi:10.1007/978-1-4757-6952-4_1

10. Murray, B. J., O'sullivan, D., Atkinson, J. D. & Webb, M. E. Ice nucleation by particles

immersed in supercooled cloud droplets. *Chem. Soc. Rev.* **41**, 6519–6554 (2012).

11. Broadley, S. L. *et al.* Immersion mode heterogeneous ice nucleation by an illite rich powder representative of atmospheric mineral dust. *Atmos. Chem. Phys.* **12**, 287–307 (2012).

12. Herbert, R. J., Murray, B. J., Whale, T. F., Dobbie, S. J. & Atkinson, J. D. Representing time-dependent freezing behaviour in immersion mode ice nucleation. *Atmos. Chem. Phys.* **14**, 8501–8520 (2014).

13. Marcolli, C., Gedamke, S., Peter, T. & Zobrist, B. Efficiency of immersion mode ice nucleation on surrogates of mineral dust. *Atmos. Chem. Phys.* **7**, 5081–5091 (2007).

14. Murray, B. J., Broadley, S. L., Wilson, T. W., Atkinson, J. D. & Wills, R. H. Heterogeneous freezing of water droplets containing kaolinite particles. *Atmos. Chem. Phys.* **11**, 4191–4207 (2011).

15. Deck, L.-T. & Mazzotti, M. Characterizing and measuring the ice nucleation kinetics of aqueous solutions in vials. *Chem. Eng. Sci.* **272**, 118531 (2023).

16. Sear, R. P. On the Interpretation of Quantitative Experimental Data on Nucleation Rates Using Classical Nucleation Theory. *J. Phys. Chem. B* **110**, 21944–21949 (2006).

17. Sear, R. P. Non-self-averaging nucleation rate due to quenched disorder. *J. Phys. Condens. Matter* **24**, 52205 (2012).

18. Sear, R. P. Statistical theory of nucleation in the presence of uncharacterized impurities. *Phys. Rev. E* **70**, 21605 (2004).

19. Holden, M. A. *et al.* High-speed imaging of ice nucleation in water proves the existence of active sites. *Sci. Adv.* **5**, (2019).

20. Campbell, J. M., Meldrum, F. C. & Christenson, H. K. Observing the formation of ice and organic crystals in active sites. *Proc. Natl. Acad. Sci. U. S. A.* **114**, 810–815 (2017).

21. Taloni, A., Vodret, M., Costantini, G. & Zapperi, S. Size effects on the fracture of microscale and nanoscale materials. *Nat. Rev. Mater.* **3**, 211–224 (2018).

22. Castillo, E. Extreme value theory in engineering. 389 (1988).

23. Levine, J. STATISTICAL EXPLANATION OF SPONTANEOUS FREEZING OF WATER DROPLETS. (1950).

24. Sear, R. P. Generalisation of Levine's prediction for the distribution of freezing temperatures of droplets: A general singular model for ice nucleation. *Atmos. Chem. Phys.* **13**, 7215–7223 (2013).

25. Langham, E. J. & Mason, B. J. The heterogeneous and homogeneous nucleation of supercooled water. *Proc. R. Soc. London. Ser. A. Math. Phys. Sci.* **247**, 493–504 (1958).

26. Bardsley, W. E. & Khatep, M. M. A General Model for Temperature of Heterogeneous Nucleation of Supercooled Water Droplets. *J. Atmos. Sci.* **41**, 856–862 (1984).

27. Sear, R. P. Quantitative studies of crystal nucleation at constant supersaturation: experimental data and models. *CrystEngComm* **16**, 6506–6522 (2014).

28. Consiglio, A. N., Rubinsky, B. & Powell-Palm, M. J. Relating Metabolism Suppression and Nucleation Probability During Supercooled Biopreservation. *J. Biomech. Eng.* (2022). doi:10.1115/1.4054217

29. Consiglio, A. N., Lilley, D., Prasher, R., Rubinsky, B. & Powell-Palm, M. J. Methods to stabilize aqueous supercooling identified by use of an isochoric nucleation detection (INDe) device. *Cryobiology* (2022). doi:10.1016/j.cryobiol.2022.03.003

30. Consiglio, A., Ukpai, G., Rubinsky, B. & Powell-Palm, M. J. Suppression of cavitation-induced nucleation in systems under isochoric confinement. *Phys. Rev. Res.* **2**, 023350 (2020).

31. Powell-Palm, M. J., Rubinsky, B. & Sun, W. Freezing water at constant volume and under confinement. *Commun. Phys.* **3**, (2020).

32. Powell-Palm, M. J., Koh-Bell, A. & Rubinsky, B. Isochoric conditions enhance stability of metastable supercooled water. *Appl. Phys. Lett.* **116**, 123702 (2020).

33. Powell-Palm, M. J. *et al.* Isochoric supercooled preservation and revival of human cardiac microtissues. *Commun. Biol. 2021 41* **4**, 1–7 (2021).

34. Koop, T., Luo, B., Biermann, U. M., Crutzen, P. J. & Peter, T. Freezing of HNO3/H2SO4/H2O Solutions at Stratospheric Temperatures: Nucleation Statistics and Experiments. *J. Phys. Chem. A* **101**, 1117–1133 (1997).

35. Kashchiev, D. *Nucleation basic theory with applications*. (Butterworth-Heinemann, 2000).

36. Zobrist, B., Koop, T., Luo, B. P., Marcolli, C. & Peter, T. Heterogeneous ice nucleation rate coefficient of water droplets coated by a nonadecanol monolayer. *J. Phys. Chem. C* **111**, 2149–2155 (2007).

37. Ickes, L., Welti, A. & Lohmann, U. Classical nucleation theory of immersion freezing: sensitivity of contact angle schemes to thermodynamic and kinetic parameters. *Atmos. Chem. Phys.* **17**, 1713–1739 (2017).

38. Hobbs, P. V. *Ice physics*. (Clarendon Press, 1974).

39. Wood, G. R. & Walton, A. G. Homogeneous Nucleation Kinetics of Ice from Water. *J. Appl. Phys.* **41**, 3027–3036 (1970).

40. Hoffman, J. D. Thermodynamic driving force in nucleation and growth processes. *The Journal of Chemical Physics* **29**, 1192–1193 (1958).

41. Wylie, R. G. The Freezing of Supercooled Water in Glass. *Proc. Phys. Soc. Sect. B* **66**, 241 (1953).

42. Diao, Y., Myerson, A. S., Hatton, T. A. & Trout, B. L. Surface Design for Controlled Crystallization: The Role of Surface Chemistry and Nanoscale Pores in Heterogeneous Nucleation. *Langmuir* **27**, 5324–5334 (2011).

43. Harrison, A. D. *et al.* Not all feldspars are equal: a survey of ice nucleating properties across the feldspar group of minerals. *Atmos. Chem. Phys.* **16**, 10927–10940 (2016).

44. Eickhoff, L. *et al.* Ice nucleation in aqueous solutions of short- and long-chain poly(vinyl

45. Gurganus, C. W., Charnawskas, J. C., Kostinski, A. B. & Shaw, R. A. Nucleation at the Contact Line Observed on Nanotextured Surfaces. *Phys. Rev. Lett.* **113**, 235701 (2014).

46. Sear, R. What do crystals nucleate on? What is the microscopic mechanism? How can we model nucleation? *MRS Bull.* **41**, 363–368 (2016).

47. De Almeida Ribeiro, I., Meister, K. & Molinero, V. HUB: A method to model and extract the distribution of ice nucleation temperatures from drop-freezing experiments. *ChemRxiv* (2023). doi:10.26434/CHEMRXIV-2022-DDZV8-V3

48. Gumbel, E. J. *Statistics of extremes*. (Dover Publications, 2004).

49. Sear, R. P. Estimation of the scaling of the nucleation time with volume when the nucleation rate does not exist. *Cryst. Growth Des.* **13**, 1329–1333 (2013).

50. Wilson, P. W. & Haymet, A. D. J. The spread of nucleation temperatures of a sample of supercooled liquid is independent of the average nucleation temperature. *J. Phys. Chem. B* **116**, 13472–13475 (2012).

51. Gumbel, E. J. & Mustafi, C. K. Some Analytical Properties of Bivariate Extremal Distributions. *J. Am. Stat. Assoc.* **62**, 569–588 (1967).

52. Balakrishna, N. & Lai, C. D. Bivariate Extreme-Value Distributions BT - Continuous Bivariate Distributions: Second Edition. in (eds. Lai, C. D. & Balakrishnan, N.) 563–590 (Springer New York, 2009). doi:10.1007/b101765_13

53. Beirlant, J., Goegebeur, Y., Teugels, J. & Segers, J. *Statistics of Extremes. Statistics of Extremes: Theory and Applications* (Wiley, 2004). doi:10.1002/0470012382

54. Vali, G. Repeatability and randomness in heterogeneous freezing nucleation. *Atmos. Chem. Phys.* **8**, 5017–5031 (2008).

55. Vali, G. Interpretation of freezing nucleation experiments: singular and stochastic; sites and surfaces. *Atmos. Chem. Phys.* **14**, 5271–5294 (2014).

56. Deubener, J. & Schmelzer, J. W. P. Statistical Approach to Crystal Nucleation in Glass-Forming Liquids. *Entropy 2021, Vol. 23, Page 246* **23**, 246 (2021).

57. Turnbull, D. Kinetics of heterogeneous nucleation. *J. Chem. Phys.* **18**, 198–203 (1950).

58. Bigg, E. K. The Supercooling of Water. *Proc. Phys. Soc. Sect. B* **66**, 688 (1953).

59. Bigg, E. K. The formation of atmospheric ice crystals by the freezing of droplets. *Q. J. R. Meteorol. Soc.* **79**, 510–519 (1953).

60. Mossop, S. C. The Freezing of Supercooled Water. *Proc. Phys. Soc. Sect. B* **68**, 193 (1955).